\documentclass [12pt,a4paper,draft]{article}
\usepackage{times}

\DeclareFontFamily{OT1}{times}{}
\DeclareFontShape {OT1}{times}{m }{n }{ <-> ptmr }{}
\DeclareFontShape {OT1}{times}{bx}{n }{ <-> ptmb }{}
\DeclareFontShape {OT1}{times}{m }{it}{ <-> ptmri}{}
\DeclareFontShape {OT1}{times}{bx}{it}{ <-> ptmbi}{}
\usepackage{amsmath}
\usepackage{amsfonts}
\begin{document}

{\LARGE {\bf \noindent On the ``equivalence'' of the Maxwell and Dirac equations}}

\vspace{1\baselineskip}

{\bf \noindent Andr\'e Gsponer}\footnote{Independent Scientific Research Institute, Box 30, CH-1211 Geneva-12, Switzerland.\\ {\bf ~~~ ~~~ isri@vtx.ch}}

\vspace{1\baselineskip}

{\noindent Document ISRI-01-07 ~~ published ~ in  ~~ \emph{ Int. J. Theor. Phys.} {\bf 41} (2002) 689--694  }

\vspace{1\baselineskip}

{\small \noindent \emph{It is shown that Maxwell's equation cannot be put into a spinor form that is equivalent to Dirac's equation. First of all, the spinor  $\pmb{\psi}$  in the representation $\vec{F} = \pmb{\psi} \vec{u} \overline{\pmb{\psi}}$ of the electromagnetic field bivector depends on only three independent complex components whereas the Dirac spinor depends on four. Second, Dirac's equation implies a complex structure specific to spin 1/2 particles that has no counterpart in Maxwell's equation.  This complex structure makes fermions essentially different from bosons and therefore insures that there is no physically meaningful way to transform Maxwell's and Dirac's equations into each other.}}

\vspace{1\baselineskip}

\vspace{1\baselineskip}
{\bf \noindent 1. INTRODUCTION}
\vspace{1\baselineskip}

The conventional view is that spin 1 and spin 1/2 particles belong to distinct irreducible representations of the Poincaré group, so that there should be no connection between the Maxwell and Dirac equations describing the dynamics of these particles.

However, it is well known that Maxwell's and Dirac's equations can be written in a number of different forms, and that in some of them these equations look very similar (e.g., Fushchich and Nikitin, 1987; Good, 1957; Kobe 1999; Moses, 1959; Rodrigues and Capelas de Oliviera, 1990; Sachs and Schwebel, 1962).  This has lead  to speculations on the possibility that these similarities could stem from a relationship that would be not merely formal but more profound (Campolattaro, 1990, 1997), or that in some sense  Maxwell's and Dirac's equations could even be ``equivalent'' (Rodrigues and Vaz, 1998; Vaz and Rodrigues, 1993, 1997).

The purpose of this paper is to investigate these possibilities and to give some arguments confirming that these formal similarities cannot lead to a physically meaningful identification of Maxwell's and Dirac's equations. To facilitate this investigation, and to make all calculations explicit, Lanczos's  biquaternionic formulation of Maxwells' and Dirac's equations will be used (Gsponer and Hurni, 1998, 2001; Lanczos, 1929).  The mathematical advantages of this formalism (which uses only complex numbers and the quaternion algebra) is that it is irreducible in the sense that, compared to formulations using larger Clifford algebras, the number of explicit components, symbols, and operations is minimal.  Moreover, like all formulations based on Clifford algebras, most calculations are in general simpler than with the standard formulations based on tensors, spinors, and matrices.

\vspace{1\baselineskip}
{\bf \noindent 2. THE ELECTROMAGNETIC FIELD AND MAXWELL'S EQUATION IN SPINOR FORM}
\vspace{1\baselineskip}

The starting point of Campolattaro's and Rodrigues's formulations of Maxwel\-l's equation is to write the electromagnetic field in spinor form using the standard $\gamma$-matrices formalism (Campolattaro, 1990)
$$
F^{\mu \nu} \equiv \overline{\Psi} S^{\mu \nu} \Psi ~~~,       \eqno{(1)}
$$
or the Clifford bundle formalism of (Vaz and Rodrigues, 1993, 1997)
$$
\mathcal{F} \equiv \psi \gamma_{21} \psi^\sim ~~~       \eqno{(2)}
$$
where the involution $(~)^\sim$ is the reversion operation in the Clifford algebra.

Maxwell's first and second equations are then written
$$
 \partial_\mu \overline{\Psi} \gamma^5 S^{\mu \nu} \Psi = 0 ~~~,~~~  \partial_\mu \overline{\Psi} S^{\mu \nu} \Psi = j^\mu ~~~,      \eqno{(3)}
$$
or, respectively,
$$
    \pmb{\partial} \psi \gamma_{21} \psi^\sim -  (\pmb{\partial}\psi \gamma_{21} \psi^\sim)^\sim = 0 ~~~,~~~ \pmb{\partial}\psi \gamma_{21} \psi^\sim +  (\pmb{\partial}\psi \gamma_{21} \psi^\sim)^\sim = 2 \mathcal{J} ~~~.      \eqno{(4)}
$$

Equations (3) and (4) are strictly equivalent to Maxwell's equations. However, the spinors $\Psi$ or $\psi$ are \underline{not} equivalent to a Dirac spinor because they have only six independent real components while the Dirac spinor has eight.  To see this explicitly, we rewrite (2)  in the biquaternion formalism
$$
   \vec{F} = \vec{E} + i\vec{B} \equiv \pmb{\psi} \vec{u} \overline{\pmb{\psi}} = \rho e^{i\beta} \mathcal{L} \vec{u} \overline{\mathcal{L}}          \eqno{(5)}
$$
where $\vec{u}$ is a constant unit vector, $\mathcal{L}$ a unit biquaternion, and the complex factor $\rho e^{i\beta}$ corresponds to a duality transformation (Rainich, 1925).  This representation is general because any non-null electromagnetic field can always be obtained by means of a duality transformation and of a Lorentz transformation $\mathcal{L} (~) \overline{\mathcal{L}}$ from a reference frame in which the electric and magnetic field vectors are parallel (Landau and Lifshitz, 1985, see Sections 24 and 25; Misner \emph{et al.}, 1970, see Exercise 20.7).

 However, expressions (1), (2) and (5) are invariant under any gauge transformation of $\Psi$,  $\psi$, or $\pmb{\psi}$ which commutes with $S^{\mu\nu}$, $\gamma_{21}$, or $\vec{u}$.  For instance, the substitution $\mathcal{L} \rightarrow \mathcal{L}\exp(c\vec{u})$ with $c \in \mathbb{C}$ leaves (5) invariant.  Therefore, the effective Lorentz transformation in (5) depends not on six but just on four parameters.  This can be seen explicitly by solving (5) for $\mathcal{L}$.  It comes
$$
     \mathcal{L} = \frac{\vec{f}+\vec{u}}{\sqrt{2(1+\vec{f}\cdot\vec{u})}}      \eqno{(6)}
$$
where $\vec{f}$ is the unit vector such that $\vec{F}=\rho e^{i\beta}\vec{f}$. This expression, which was first derived by Gürsey (1956, p.~167), confirms that the spinor $\pmb{\psi}=\sqrt{\rho} e^{i\beta/2}\mathcal{L}$ associated with an electromagnetic field has three complex components, and not four like a Dirac spinor.

\vspace{1\baselineskip}
{\bf \noindent 3. MAXWELL'S EQUATION IN DIRAC-LIKE FORM}
\vspace{1\baselineskip}

By a number of lengthy tensor manipulations Campolattaro (1990) succeeded in reducing equations (2) to a single nonlinear equation in which the four-gradient of the spinor $\Psi$ appears on one side, and in which the nonlinearity appears on the other side as a complicated variable factor that would be the mass if $\Psi$ was a Dirac field.  The same calculation was repeated by Vaz and Rodrigues (1993) who confirmed the power of the Clifford number formalism by deriving an equivalent equation in a very straightforward manner. In the formalism of Rodrigues (1997, 1998), Maxwell's equations (4) are equivalent to the nonlinear equation
$$
\pmb{\partial} \psi \gamma_{21} = \frac{\exp(\gamma_5\beta)}{\rho} \left[\frac{1}{2}\mathcal{J} +(j + \gamma_5 g)   \right] \psi       \eqno{(7)}
$$
where $j=\gamma^\mu\langle(\partial_\mu\psi)\gamma_{21}\psi^\sim\rangle$ and $g=\gamma^\mu\langle(\partial_\mu\psi)\gamma_5\gamma_{21}\psi^\sim\rangle$.  In the case where $\mathcal{J}=0$, and provided that $\rho$, $\beta$,  $j$ and $g$ are constants, this expression would be similar to Dirac's equation, which in Rodrigues's formalism is 
$$
\pmb{\partial} \psi_D \gamma_{21} = m ~ \psi_D  ~~~.     \eqno{(8)}
$$
However, since  $\psi$ in (7) has only six real functions in its components, it cannot be made equivalent to (8) in the general case where the Dirac spinor $\psi_D$ has eight independent components.  For the same reason, contrary to what Campolattaro and Rodrigues tried to do, it is not possible to find non-trivial cases in which the mass term in Maxwell's equation in spinor form becomes a constant. In particular, it is not possible to use constraints such as $\pmb{\partial} \cdot j = 0$ and $\pmb{\partial} \cdot g = 0$ because they reduce the number of independent real components from six to four.

\vspace{1\baselineskip}
{\bf \noindent 4. INTRINSIC DIFFERENCE BETWEEN MAXWELL'S AND DI\-RAC'S EQUATIONS}
\vspace{1\baselineskip}

Let us translate equations (4) and (8) into the biquaternion formalism (Gsponer and Hurni, 1998, 2001). Max\-well's equations are then
$$
    \nabla \vec{F} - \vec{F}^\sim \nabla = 0  ~~~,~~~ \nabla \vec{F} + \vec{F}^\sim \nabla = 2 J ~~~, \eqno(9)
$$
 and Dirac's equation becomes the \emph{Dirac-Lanczos equation} (Lanczos, 1929)
$$
    \overline{\nabla} \psi_D = i m ~ \psi_D^* ~ \vec{u}   ~~~.   \eqno(10)    
$$
Both equations are of first order, which is why the question of their possible ``equivalence'' can arise.  However, there is one essential difference: Dirac's equation relates the field $\psi_D$ to its complex conjugate $\psi_D^*$, while Maxwell's equation (as well as Proca's equation for a massive spin 1 particle) do not. This complex conjugation operation  arises naturally when the spin 1/2 and spin 1 field equations are consistently derived from Lanczos's fundamental equation (Gsponer and Hurni, 1998, 2001), and it is intimately connected to the fact that fermions are essentially different from bosons (Gsponer and Hurni, 1994). Indeed, by studying the time-reversal transformation of (10), one finds $T^2 = -1$, which implies the Pauli exclusion principle and Fermi statistics (Feynman, 1987; Weinberg, 1998, p.~80). 

Therefore, since the spinor representation of the electromagnetic field (5) involves $\pmb{\psi}$ and $\overline{\pmb{\psi}}$, not $\pmb{\psi}$ and $\overline{\pmb{\psi}^*}$, it is impossible to transform Maxwell's equations (9) by any algebraic manipulation to get a Dirac equation for the spinor $\pmb{\psi}$, even if we accept a nonlinear variable mass term.

The complex conjugation operation that is explicit in (10) has a counterpart in any formulation of Dirac's equation.  Its absence in the standard $\gamma$-matrices formulation or in other Clifford formulations of Dirac's equation, e.g. (8), needs therefore to be explained.  The reason is that  these formulations either use a larger algebra, so that the complex conjugation in (10) can be avoided, or a ``complex structure'' (or ``complex geometry'') by which operations on complex numbers are replaced by algebraic operations on real numbers (De Leo, 2001; This is a recent example in which ordinary complex numbers are replaced by linear functions of real quaternions).

For example, in the standard  $\gamma$-matrices formulation the Dirac algebra $\mathbb{D} \sim \mathcal{M}_4(\mathbb{C})$ is used instead of the biquaternion algebra $\mathbb{B} \sim \mathcal{M}_2(\mathbb{C})$. Then, the complex conjugate of the Dirac matrices is given by ${\gamma_\mu}^* = \gamma_0 {\gamma_\mu}^t \gamma_0$ where $(~)^t$ denotes transposition, and the ``Dirac adjoint'' $\overline{\Psi} \equiv \Psi^\dagger \gamma_0$ is required instead of $\Psi^\dagger$ in scalar products (see, e.g., Weinberg, 1998). Similarly, Lanczos's ``complex'' formulation of Dirac's equation (10) can be transformed into the ``real'' formulation of Hestenes (1966, see sections 7 and 13) by using the identification $D^*  = \gamma_0 D \gamma_0$ to replace the symbolic complex conjugation operator $(~)^*$ by the linear function $\gamma_0 (~) \gamma_0$ (see, in particular, Fauser, 2001, where it is shown that this function provides a link between seemingly inequivalent formulations of Dirac's equation).

\vspace{1\baselineskip}
{\bf \noindent 5. CONCLUSION}
\vspace{1\baselineskip}

By various algebraic transformations it is possible to put Maxwell's equation into a form that is very similar to Dirac's equation.  Similarly, it is also possible to replace the complex conjugation operator that is explicit in the Dirac-Lanczos equation (10) by an algebraic operation that effectively implements this complex conjugation.  However, the transformed Maxwell equation will still depend on only six real functions while  Dirac's equation requires eight, and the complex structure which makes a Dirac particle a fermion rather than a boson will remain an essential feature of Dirac's equation.

Indeed, the complex structure inherent to Dirac's equation corresponds to the fact that the Dirac field is only meaningful in the context of quantum mechanics where complex numbers are essential. On the other hand, such a complex structure is absent in Maxwell's equation so that Maxwell's field has a consistent interpretation in the context of classical electrodynamics where complex numbers are not essential.

There are of course other fundamental differences between the Maxwell and Dirac equations and their interpretations that are not discussed in this paper. See, e.g., (Good, 1957).

Finally, while practical calculations are made most easily using a formulation such as the standard $\gamma$-matrices, it is important to stress that the Dirac-Lanczos equation (10) has the considerable didactical advantage to make manifest the fermionic complex structure that is hidden in Dirac's original formulation, as well as in most other ones, e.g., Rodrigues's (8), or Hestenes's (Hestenes, 1966).

\vspace{1\baselineskip}
{\bf \noindent ACKNOWLEDGMENTS}
\vspace{1\baselineskip}

The author is indebted to Prof.\ W.\ A.\ Rodrigues Jr.\ and to Dr.\ J.\ Vaz for extensive correspondence, to Prof.\ W.\ E.\ Baylis for several useful remarks, and to Dr.\ J.-P.\ Hurni for illuminating discussions.
\vspace{1\baselineskip}

\vspace{1\baselineskip}
{\bf \noindent REFERENCES} 
\vspace{1\baselineskip}

\noindent Campolattaro, A. A. (1990). Generalized Maxwell equations and quantum mechanics. \emph{International Journal of Theoretical Physics} {\bf 29}, 141--155, 477--482.

\noindent Campolattaro, A. A. (1997). From classical electrodynamics to relativistic quantum mechanics. In \emph{The Theory of the Electron}, J. Keller and Z. Oziewicz, eds. \emph{Advances  in Applied Clifford Algebras}, Universidad Nacional Aut\'onoma de M\'exico, Mexico, {\bf 7 (S)}, 167--173.

\noindent  De Leo, S. (2001). Quaternionic Lorentz group and Dirac equation. \emph{Foundations of Physics Letters} {\bf 14}, 37--50.

\noindent Fauser, B. (2001). Equivalence of Daviau's, Hestenes', and Parra's formulations of Dirac theory. \emph{International Journal of Theoretical Physics} {\bf 40}, 399--411.

\noindent Fushchich, W. I., and  Nikitin, A. G. (1987). \emph{Symmetries of Maxwell's Equations}, Reidel, Dordrecht.

\noindent Feynman, R. P. (1987). The reasons for antiparticles. In   \emph{Elementary Particles and the Laws of Physics},  R. P. Feynman and S. Weinberg, eds., Cambridge University Press, New York, pp.~1--59.

\noindent Good, R. H. (1957). Particle aspect of the electromagnetic field equations. \emph{Physical Review} {\bf 105}, 1914--1919.

\noindent Gsponer, A., and Hurni J.-P. (1994). Lanczos's equation to replace Dirac's equation~? In \emph{ Proceedings of the Cornelius Lanczos International Centenary Conference},  J.D. Brown, M.T. Chu, D.C. Ellison, and R.J. Plemmons, eds., SIAM, Philadelphia, pp.~509--512.  There are a number of typographical errors in this paper. Please ask the authors for a corrected version. 

\noindent Gsponer, A., and Hurni J.-P. (1998). Lanczos-Einstein-Petiau: From Dirac's equation to non-linear wave mechanics. In \emph{Cornelius Lanczos Collected Published Papers With Commentaries}, W. R. Davis, general editor, M. T. Chu, P. Dolan, L. R. McConnell, L. K. Norris, E. Ortiz, R. J. Plemmons, D. Ridgeway, B. K. P. Scaife, W. J. Stewart, J. W. York, Jr., W. O. Doggett, B. M. Gellai, A. Gsponer, C. A. Prioli, eds., North Carolina State University, Raleigh, {\bf III}, 2-1248 -- 2-1277.

\noindent Gsponer, A., and Hurni J.-P. (2001). Comment on formulating and generalizing Dirac's, Proca's, and Maxwell's equations with biquaternions or Clifford numbers. \emph{Foundations of Physics Letters} {\bf 14}, 77--85.

\noindent Gürsey, F. (1956). Contribution to the quaternion formalism in special relativity. \emph{Revue de la Facult\'e des Sciences}, University of Istanbul, {\bf A 20}, 149--171.

\noindent Hestenes, D. (1966). \emph{Space-Time Algebra}, Gordon and Breach, New York.

\noindent Kobe, D. H. (1999). A relativistic Schr\"odinger-like equation for a photon and its second quantization. \emph{Foundations of Physics} {\bf 29}, 1203-1231. Idem. Zitterbewegung of a photon. \emph{Physics Letters} {\bf A 253}, 7--11.

\noindent Lanczos, C. (1929). Die tensoranalytischen Beziehungen der Diracschen Gleichung. \emph{Zeitschrift f\"ur Physik} {\bf 57}, 447--493. Reprinted and translated in \emph{Cornelius Lanczos Collected Published Papers With Commentaries}, W. R. Davis, general editor, M. T. Chu, P. Dolan, L. R. McConnell, L. K. Norris, E. Ortiz, R. J. Plemmons, D. Ridgeway, B. K. P. Scaife, W. J. Stewart, J. W. York, Jr., W. O. Doggett, B. M. Gellai, A. Gsponer, C. A. Prioli, eds., North Carolina State University, Raleigh, {\bf III}, 2-1132 -- 2-1225.

\noindent Landau, L. D., and Lifshitz, E.M. (1975). \emph{The Classical theory of Fields}, Pergamon Press, Oxford.

\noindent Misner, C. W., Thorn, K.S., and Wheeler, J.A. (1970). \emph{Gravitation}, Freeman and Co., San Francisco).

\noindent Moses, H. E. (1959). Solution of Maxwell's equations in spinor notation. \emph{Physical Review} {\bf 113}, 1670--1679.

\noindent Rainich, G. Y. (1925). Electrodynamics in the general relativity theory. \emph{American Mathematical Society Transactions} {\bf 27}, 106--136.

\noindent Rodrigues, W. A., Jr., and Capelas de Oliviera, E. (1990). Dirac and Maxwell equations in the Clifford and spin-Clifford bundles. \emph{International Journal of Theoretical Physics} {\bf 29}, 397--412.

\noindent Rodrigues, W. A., Jr., and Vaz, J. (1998). From electromagnetism to relativistic quantum mechanics. \emph{Foundations of Physics} {\bf 28}, 789--814.

\noindent Sachs, M., and Schwebel, S. L. (1962). On covariant formulations of the Maxwell-Lorentz theory of electromagnetism. \emph{Journal of Mathematical Physics} {\bf 3}, (1962) 843--848.

\noindent Vaz, J., and Rodrigues, W. A., Jr. (1993). Equivalence of Dirac and Maxwell equations and quantum mechanics. \emph{International Journal of Theoretical Physics} {\bf 32}, 945--959.

\noindent Vaz, J., and Rodrigues, W. A., Jr. (1997). Maxwell and Dirac theories as an already unified theory. In \emph{The Theory of the Electron}, J. Keller and Z. Oziewicz, eds. \emph{Advances  in Applied Clifford Algebras}, Universidad Nacional Aut\'onoma de M\'exico, Mexico, {\bf 7 (S)}, 369--385.

\noindent Weinberg, S. (1998). \emph{The Quantum Theory of Fields, Vol. I}, Cambridge University Press, New York.

\end{document}